\documentclass{article}
\usepackage{notosaarticle} 

\usepackage{subfig}
\usepackage[export]{adjustbox}
\usepackage{booktabs}
\usepackage{float}
\usepackage{algorithm}
\usepackage{algorithmic}

\usepackage{mathtools}
\DeclarePairedDelimiter{\floor}{\lfloor}{\rfloor}

\usepackage{tikz}
\usetikzlibrary{decorations.pathreplacing,calc}
\newcommand{\tikzmark}[1]{\tikz[overlay,remember picture] \node (#1) {};}

\newcommand*{\AddNote}[4]{%
    \begin{tikzpicture}[overlay, remember picture]
        \draw [decoration={brace,amplitude=0.5em},decorate,ultra thick,black]
            ($(#3)!(#1.north)!($(#3)-(0,1)$)$) --  
            ($(#3)!(#2.south)!($(#3)-(0,1)$)$)
                node [align=center, text width=2.5cm, pos=0.5, anchor=west] {#4};
    \end{tikzpicture}
}

\articletype{arXiv Article}

\usepackage{todonotes}

\begin{document}

\title{Blind multi-frame deconvolution for the correction of space-variant blur in images}

\author{Wouter van de Ketterij,\authormark{1,*} Oleg Soloviev,\authormark{1,2,3} and Michel Verhaegen\authormark{1}}

\address{\authormark{1}Delft Center for Systems and Control, TU Delft, Mekelweg 2, 2628 CD Delft, The Netherlands\\
\authormark{2}ITMO University, Kronverksky 49, 197101 St Petersburg, Russia\\
\authormark{3}Flexible Optical B.V., Polakweg 10-11, 2288 GG Rijswijk, The Netherlands}

\email{\authormark{*}wvandeketterij@gmail.com} 



\begin{abstract}
This paper demonstrates a practical method that can correct spatial varying blur from a set of images of the same object. The algorithm jointly estimates the object and local point spread functions~(PSF). The method prioritizes sections with small spatial variation in the PSF for deconvolution. This novel approach can handle large translations in the local PSFs, hence the algorithm is able to correct for morph in the images. Robustness to noise is demonstrated in numerical simulations. Numerical experiments are conducted where the performance of the algorithm is compared to a state-of-the-art method found in literature. The algorithm can be used in situation with space-temporal variation of the PSF and can be applied in situations where the signal-to-noise ratio is low.
\end{abstract}

\section{Introduction}
Restoring the image quality degraded by  blur and noise is an active research topic with applications in astronomy~\cite{roddier1999}, microscopy~\cite{Booth5788}, medical imaging~\cite{FOSTER}, machine vision~\cite{sonka2014}, \emph{etc}.
The blur can be caused by the optical aberration due to light propagation through inhomogeneous medium, by imperfection of the optics of imaging system, by the motion of the camera and/or the object.
In the simplest cases, the effect of the blur can be represented by convolution of the object with a spatially invariant PSF, and a number of approaches dealing with isoplanatic deconvolution with or without \emph{a priori} knowledge of PSF from one or multiple images is known~\cite{Cannon,AyersDainty,Schulz:93,Yaroslavsky:94,Wilding2017}.
In the anisoplanatic case, \emph{e.g.} in deep microscopy imaging~\cite{NaJi} or images deteriorated by motion blur~\cite{Dai2008,Hirsch2010,Sroubek2016}, the shape of the PSF is not constant over the image, the isoplanatic algorithms fail, and anisoplanatic deconvolution algorithms able to evaluate the PSF at every point of the image are required.

However, it is computationally inefficient to implement a method that calculates a unique correction for every point~\cite{Seitz2009,Hirsch2010,Denis2011}. A group of methods uses a global decomposition of the spatially varying~(SV) PSF in order to work with a limited number of spatially static PSFs. Flicker and Rigaut~\cite{Flicker2005} approximate the SV PSF by decomposition into a number of spatially invariant modes. Miraut and Portilla~\cite{Miraut2012} and later Sroubek \textit{et al.}~\cite{Sroubek2016} decompose the the SV PSF into spatially invariant modes by means of the singular value decomposition. An advantage of decomposition of the SV PSF by the singular value decomposition is that it is globally optimal with respect to the approximation error. On the other hand, the use of global decomposition methods comes with large computational load~\cite{Denis2015}.

Another approach to limit the SV PSF to a number of spatially static PSFs is division of the image into isoplanatic subsections. This method was first proposed by Trussel and Hunt~\cite{TrusselHunt}; they divided the image into disjoint subsections with local spatially static PSFs. This method was later used by Costello and Mikhael~\cite{CostelloMikhael} in combination with modified Wiener filtering in order to correct for anisoplanatic aberrations. The implementation of disjoint isoplanatic subsections is known to lead to the appearance of boundary artifacts~\cite{Denis2015}. Nagy and O'Leary~\cite{Nagy97} implemented the overlap-add method, where overlapping isoplanatic subsections were used for deconvolution. The use of this overlap-add method reduces the effects of boundary artifacts in the restored image. This method was later improved by Denis \textit{et al.}~\cite{Denis2015}.

A popular trend in image reconstruction is the use of neural networks. Shajkofci and Liebling~\cite{Shajkofci2018} used a Convolutional Neural Network that requires an optical parametric model. The optical model is used as prior knowledge of the PSF. Another approach is the use image specific \textit{a priori}. Sureau \textit{et al.}~\cite{sureau2019} used a Deep Neural Network architecture to learn supervised setting parameters adapted for galaxy image processing.

The problem becomes even more challenging when no prior knowledge of the PSF or object is available, as is generally the case with medium induced aberrations in microscopy. Hirsch \textit{et al.}~\cite{Hirsch2010} used a similar overlap-add method to perform blind anisoplanatic deconvolution with multiple images of the same object. In blind deconvolution it is the goal to estimate both the object and the aberration or (local) PSF. Many of the blind anisoplanatic correction methods found in literature~\cite{TrusselHunt,LagendijkBiemond1991,GuoLeeTeo1997} use the assumption that the image can be divided into isoplanatic subsections. This group of methods divides the image(s) into isoplanatic subsections and apply conventional isoplanatic blind deconvolution methods in order to estimate the object locally. The performance of methods in this group depends on the speed at which the SV PSF is changing in the image. The assumption that the PSF is spatially static on a small subsection does not always hold. We propose weighted multi-frame deconvolution, a method that prioritizes subsections with highly isoplanatic aberrations for deconvolution. A small weight is assigned when large variation occurs in the PSF locally and parts with small PSF variation are assigned a large weight. These weights determine the contribution to deconvolution of an image frame.

Using multiple images of the same object with different aberrations comes with another problem. The difficulty is that translations in the PSFs cause morph of the images, \textit{i.e.} parts of the object appear at different distances in the images. For local multi-frame deconvolution it is required that the isoplanatic subsections in the images correspond to the same part of the object. This paper demonstrates an algorithm, Tangential Iterative Projections Algorithm for Anisoplanatic Aberrations with Adaptive finite support and Weighted multi-frame deconvolution (TIP4AW), that is specifically designed to deal with this problem.

TIP4AW processes multiple images of the same object in batch mode. All images are considered at once in an iterative process of four steps. In the first step the object is locally estimated by weighted multi-frame deconvolution and the overlap-add method is use to form the estimated object. The second step is the projection of the object on the feasible set of all possible objects. In the third step all local PSFs are estimated by single-frame deconvolution. Isolation of the corresponding subsection in this step is performed by multiplication with an apodization kernel. The fourth and last step is the projection of the estimated local PSFs on the feasible set of all possible PSFs, using adaptive PSF support. Numerical simulations are conducted to show the robustness to noise of this algorithm. Moreover the performance of the algorithm is compared to a state-of-the art algorithm found in literature.

\section{Local blind multi-frame deconvolution}
As described above, a common approach to blind anisoplanatic deconvolution is the use of isoplanatic subsections. On isoplanatic subsections conventional blind deconvolution is performed and the local results are combined to form a corrected image. This section describes anisoplanatic and isoplanatic image formation and elaborates on deconvolution by linear filtering. Furthermore, it is explained how isoplanatic deconvolution can be used in order to deconvolve the image from anisoplanatic aberrations using overlap-add.

\subsection{Image formation}
In an incoherent imaging system the formation of an image can be seen as the the summation of the contribution of all point sources in the object, $o_{k,l}$, where $k,l$ are the discrete spatial coordinates in the object plain. Discrete coordinates are used, because an image is in general spatially samples by a CCD. Let $\tilde{h}_{m,n,k,l}$ be the normalized contribution, or PSF, of point source $o_{k,l}$, where $m,n$ are the spatial coordinates in the image plain. The formation of an image, $i$, is given by
\begin{equation}
    i_{m,n} = \sum_{k=1}^M\sum_{l=1}^N o_{k,l} \tilde{h}_{m,n,k,l} + e_{n,m}, \qquad m=1,...,M, \quad n=1,...,N ,
    \label{eq:DiscAnisoImageAndnoise}
\end{equation}
where $e_{n,m}$ is measurement noise. 

The image model of Eq.~\eqref{eq:DiscAnisoImageAndnoise} describes anisoplanatic image formation where all PSFs can be different. In the case of (almost) spatially invariant PSF, the isoplanatic imaging model is used:
\begin{equation}
    i_{m,n} = \sum_{k=1}^M\sum_{l=1}^N o_{k,l} h_{m-k,n-l} + e_{m,n},
    \label{eq:DiscIsoImageAndnoise}
\end{equation}
Equation~\eqref{eq:DiscIsoImageAndnoise} can be recognized as convolution of the object and the PSF, and is denoted in short notation by $i_{m,n}=o_{m,n}\ast h_{m,n} + e_{m,n}$. For the isoplanatic model the convolution theorem can be used and in the spatial frequency domain the convolution in Eq.~\eqref{eq:DiscIsoImageAndnoise} becomes the point-wise multiplication,
\begin{equation}
    I_{m,n} = O_{m,n}\times H_{m,n} + E_{m,n},
    \label{eq:OTF}
\end{equation}
where the capital letter describes a two dimensional Discrete Fourier Transform. The PSF spectrum, $H_{m,n}$, in Eq.~\eqref{eq:OTF} is referred to as the optical transfer function ~(OTF).

\subsection{Deconvolution}
Assuming that Eqs.~\eqref{eq:DiscIsoImageAndnoise} and \eqref{eq:OTF} hold and that the PSF is known, a na\"{i}ve approach to deconvolution is point-wise division of the image spectrum and OTF, $\hat{O}_{m,n}=I_{m,n}/H_{m,n}$. This approach, however, is known to amplify noise and usually leads to a decrease in the signal-to-noise ratio (SNR)~\cite{DecLab2}. A more optimal method is given by the Wiener filter~\cite{Helstrom67},
\begin{equation}
    \hat{O}_{m,n} = \frac{(H_{m,n})^*I_{m,n}}{|H_{m,n}|^2+\frac{1}{\text{SNR}}},
    \label{eq:Wiener}
\end{equation}
where $H^*$ denotes the complex conjugate of $H$. 
Equation~\eqref{eq:Wiener} was later modified by Gonzalez and Woods~\cite{GonWoods} for situations where the SNR is unknown.

\subsection{Blind multi-frame deconvolution}
In some applications the exact shape of the PSF is not known, and both the object and PSF should be estimated. Blind deconvolution can be posed as an ill-conditioned minimization problem~\cite{Chaudhuri}. Ayers and Dainty~\cite{AyersDainty} were one of the first to propose a practical algorithm that can solve this problem by alternating between solutions for the object and the PSF and applying constraints in the Fourier domain and spatial domain. Other blind deconvolution methods make use of the temporal variation of the PSF. More than one image of the same object, but with different aberration are used to reconstruct the object and PSF. Such methods are referred to as blind multi-frame deconvolution (BMFD) or blind multi-channel deconvolution. In BMFD a set of $S$ images $\{i_{m,n|s}\}$, with corresponding unknown PSFs $\{h_{m,n|s}\}$, are deconvolved. All images are obtained from the same object,
\begin{equation}
    i_{m,n|s} = \sum_{k=1}^M\sum_{l=1}^N o_{k,l} h_{m-k,n-l|s} + e_{m,n|s}.
    \label{eq:Multi-FrameFormation}
\end{equation}
Different images from the same object can be obtained if the aberration is time-variant. This is often the case in \textit{in vivo} microscopy~\cite{Wilding2018} and in imaging trough turbulence~\cite{Roggemann95}. Schulz~\cite{Schulz:93} demonstrated in 1993 a practical implementation of an algorithm that uses more than one image. Schulz's algorithm is iterative in nature and uses penalized maximum likelihood estimation and a parameterization  of the PSF in order to jointly estimate the object and PSF. A year later, Yaroslavsky \textit{et al.}~\cite{Yaroslavsky:94} showed that a Wiener-like filter,
\begin{equation}
    \hat{O}_{m,n} = 
    \frac{\sum_{s=1}^S (H_{m,n|s})^*I_{m,n|s}}{\sum_{s=1}^S |H_{m,n|s}|^2},
    \label{eq:Multi-Frame-Wiener}
\end{equation} 
can be used to estimate the object from more than one image when the PSFs are known. In 2017, Wilding \textit{et al.}~\cite{Wilding2017} demonstrate an Ayers and Dainty-like algorithm that uses the filter in Eq.~\eqref{eq:Multi-Frame-Wiener} in order to estimate the object from multiple images. Equation~\eqref{eq:Multi-Frame-Wiener} fails at locations where common zeros in the PSF spectrum occur. Therefore, it is best to record as much images with different PSFs as possible. Since common zeros in the PSF spectrum are more likely to occur in one dimensional~(1D) signals, this algorithm is not optimal in a 1D analysis.

\subsection{Spatial variation of the PSF}
The assumption that the PSF is spatially static is a necessary condition for Eq.~\eqref{eq:OTF} to hold. For anisoplanatic aberrations it can be assumed that Eq.~\eqref{eq:OTF} holds for small enough regions in the image, isoplanatic subsections. This allows the conventional deconvolution techniques to be applied locally. Local results are combined to form the estimated object.

Trussel and Hunt~\cite{TrusselHunt} combined the local results by adding disjoint isoplanatic sections together. This method results in boundary artefacts near the edges of isoplanatic subsections. Nagy and O'Leary~\cite{Nagy97} used the overlap-add method where the isoplanatic subsections overlap each other. An isoplanatic subsection was defined by windows, $W_{\tilde{p}}[m,n]$, where the index $\tilde{p}$ is the window number and $m,n$ are discrete spatial coordinates of the window. These windows must have the following property:
\begin{equation}
    \sum_{\tilde{p}=1}^{\tilde{p}_t} W_{\tilde{p}}[m,n] = 1,
\end{equation}
where $\tilde{p}_t$ is the total number of windows. In overlap-add, the image is locally (de)convolved with a local PSF by means of applying a window $W_{\tilde{p}}$. The global result is found after summation of local results.

\section{Algorithm description}
TIP4AW uses tangential iterative projections~\cite{Wilding2017} and the overlap-add method~\cite{Nagy97,Denis2015} in conjunction with novelties weighted multi-frame deconvolution, Section \ref{:ObjectFourier}, and adaptive PSF support, Section \ref{:PSFSpatial}, to handle anisoplanatic aberrations in multiple images of the same object. It is an algorithm that is demonstrated to be robust to noise and morph.

The method to correct for anisoplanatic aberrations in blurred images requires multiple images of the same object. These images are divided into overlapping subsections such that overlap-add can be applied after deconvolution to avoid the appearance of boundary artifacts. Let $S$ be the number of available images, and let $i_{p,q|s}[m,n]$ denote subsection $p,q$ of image $i_s$. The discrete spatial coordinates $m,n$ are not denoted anymore for ease of reading. Section~\ref{::Discussion} provides more details on the relation between the number of subsections and the quality of the estimated object. Figure~\ref{fig:ImageSubsections}(a) shows an image that is divided into subsections. The subsection $i_{p,q}$ is shaded in red. Every subsection has four quadrants. Subsections $i_{p,q}$ and $i_{p,q+1}$ have two overlapping quadrants. The local PSF for subsection $i_{p,q}$ in image $s$ is denoted by $h_{p,q|s}$. TIP4AW proceeds by iteratively improving estimates of the object and PSF.

Each iteration consists of four steps. In the first step weighted multi-frame deconvolution is used in order to estimate the object locally. Overlap-add is then used to combine local results and find the estimated object. In the second step, the estimated object is improved by applying spatial domain constraints. During the third step, the local PSF is estimated after applying local apodization on the object. In the forth and final step, PSF constraints in the spatial domain are applied to improve the estimated local PSFs.

\subsection{Object step in Fourier domain}
\label{:ObjectFourier}
\begin{figure}[!ht]%
\centering
\includegraphics[width=\textwidth]{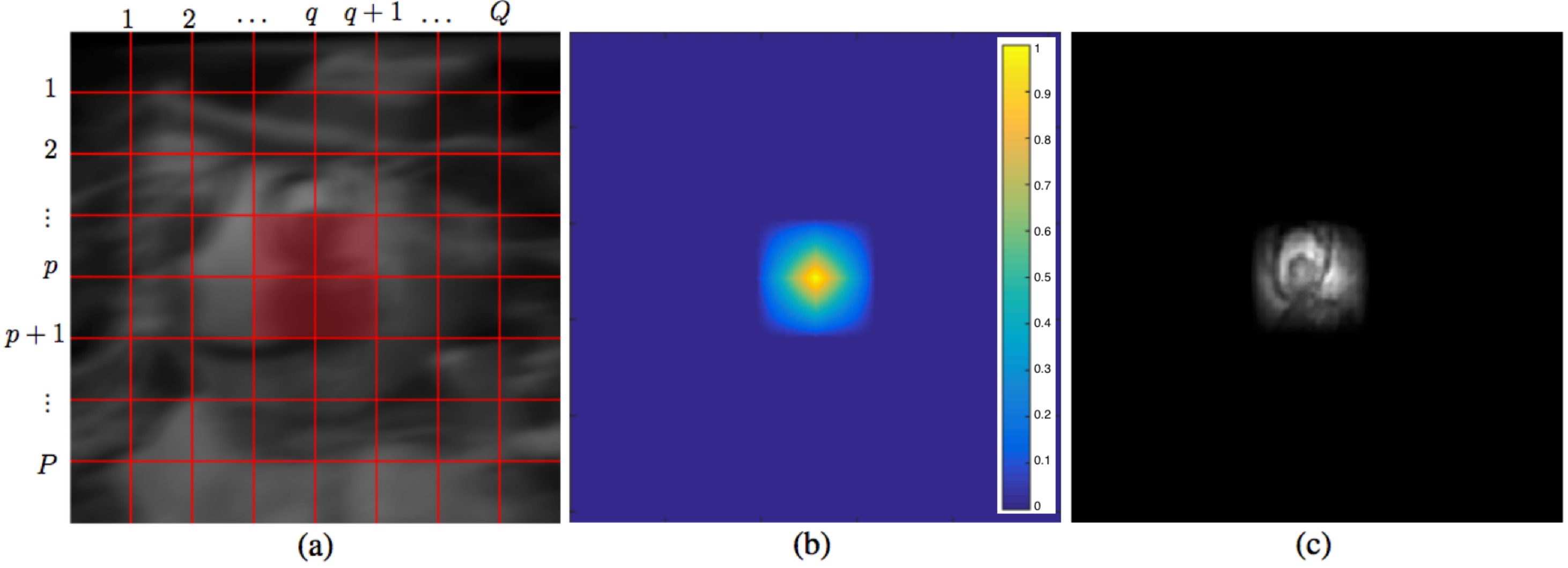}
    \caption{(a)~The deteriorated image is divided into $P$ by $Q$ overlapping subsections. Subsection $i_{p,q}$ is highlighted in red. Every subsection has four quadrants. (b)~The bi-linear interpolation masks, Eq.~\ref{eq:mask}, are used to create one object from all overlapping subsections. Every local object is multiplied with a corresponding mask or window. The sum of these masks is one uniformly. The values of these masks represent the contribution to a point as in bi-linear interpolation. (c)~The object subsection is the local object multiplied with the interpolation mask. The estimated object is the sum of all object subsections.}%
    \label{fig:ImageSubsections}%
\end{figure}

The algorithm is initialized by assuming that all local PSFs are delta functions, $h_{p,q|s} = \delta[\floor*{m/2},\floor*{n/2}]$, where $\floor*{\cdot}$ is the floor function. As a result, the first estimated object is a pixel-wise average of all images. During the $k$th iteration, weighted multi-frame deconvolution is used to estimate the local object
\begin{equation}
    \bar{o}_{p,q}^{(k)} = \mathcal{F}^{-1} \left\{ \frac{\sum_{s=1}^S a^{(k)}_{p,q|s} \left(H_{p,q|s}^{(k)}\right)^*I_{s}}{\left[ \sum_{s=1}^S a^{(k)}_{p,q|s} |H_{p,q|s}^{(k)}|^2\right]_{>\epsilon\bar{a}_{p,q}}} \right\},
    \label{eq:WMFD}
\end{equation}
where $\mathcal{F}^{-1}\{\cdot\}$ denotes the inverse Discrete Fourier Transform and multiplication and division are element-wise. The notation $N/[D]_{>\alpha}$ means that division is performed if the absolute value of the denominator is larger than $\alpha$, else the denominator is set to infinity, such that the result becomes zero. In Section \ref{sec:weights} is explained how the weights, $a^{(k)}_{p,q|s}$, are determined. $\bar{a}_{p,q}$ is the average $\frac{1}{S}\sum_{s=1}^Sa^{(k)}_{p,q|s}$. The parameter $\epsilon>0$ is related to the SNR, if the SNR is unknown, a small $\epsilon$ can be chosen. Every local object is multiplied element-wise by a window, $W_{p,q}$. The windows are chosen such that $\sum_{p=1}^P\sum_{q=1}^Q W_{p,q}=1$ for every pixel. Here, the windows are congruent to a bi-linear interpolation mask:
\begin{equation}
   W_{p,q}[m,n] = \max\left\{1-\frac{|c_p-m|}{l_m/2},0 \right\}\cdot\max\left\{1-\frac{|c_q-n|}{l_n/2},0\right\},
   \label{eq:mask}
\end{equation}
where $C[c_p,c_q]$ is the centre of subsection $p,q$ and $l_m$ and $l_n$ are the vertical and horizontal lengths of the subsection respectively, and $\max\{\cdot\}$ is a pixel-wise operation. Figure~\ref{fig:ImageSubsections}(b) shows the bi-linear interpolation mask in subsection $p,q$. Extra care must be taken when dealing with exterior subsections, subsections where $p$ or $q$ is equal to $1,P$ or $Q$. All interior subsections are overlapped by four neighbouring subsections and therefore the sum of bi-linear interpolations masks is one uniformly. This is, however, not the case for exterior subsections. It is possible to use local PSFs on the edge of the images. However, it is more difficult to calculate local PSFs on the edge. Instead of calculating the local PSF on the edge, the closest interior PSF is chosen. This is similar to normalizing the windows in Eq.~\eqref{eq:mask}, $W_{p,q}[m,n] = {W_{p,q}[m,n]}/{\sum_{p=1}^P\sum_{q=1}^QW_{p,q}[m,n]}$.
After multiplication of the local object with the corresponding window, the object subsection is formed, Fig.~\ref{fig:ImageSubsections}(c). Summation of these object subsection creates the estimated object
\begin{equation}
    \bar{o}^{(k)} = \sum_{p=1}^P\sum_{q=1}^Q \bar{o}^{(k)}_{p,q}W_{p,q}.
    \label{eq:objectsynthesis}
\end{equation}

\subsection{Object step in spatial domain}
\label{:ObjectSpatial}
In the second step, the object is projected on the feasible set of all possible object. This is done by applying the following object constraints. The intensity of the estimated object must be non-negative and the object is normalized such that $\left|\left|o^{(k)}\right|\right|_{1} = 1$.
\begin{equation}
    o^{(k)} = \frac{\max\left\{ \text{real}\left(\bar{o}^{(k)}\right),0 \right\}}{\left|\left|\max\left\{ \text{real}\left(\bar{o}^{(k)}\right) ,0\right\}\right|\right|_{1}}.
\end{equation}

\subsection{PSF step in Fourier domain}
\label{:PSFFourier}
In this step, the local PSFs are estimated by local single-frame deconvolution. The image can be deconvolved with the object locally by isolating the correct part in the object. A Gaussian apodization kernel is used to isolate the corresponding part of the object. Apodization also reduces edge effects, such as ringing and ghosts~\cite{DecLab2}. Apodization centred in subsection $p,q$ with center $C[c_p,c_q]$ is performed by multiplication of the object with kernel
\begin{equation}
    K^{w}_{p,q}[m,n] = e^{-\frac{(m-c_p)^2+(n-c_q)^2}{w^2}},
\end{equation}
where $w$ is the width parameter of the Gaussian apodization kernel. $K^{w}_{p,q}$ is the apodization kernel corresponding to local object, $o_{p,q}$, and local PSFs, $h_{p,q|s}$. Then the estimated local PSF is given by single-frame deconvolution of the image and apodized object.
\begin{equation}
   \bar{h}_{p,q|s}^{(k+1)} = \mathcal{F}^{-1} \left\{  \frac{I_s}{ \left[ O_{K^{w}_{p,q}}^{(k)} \right]_{>\epsilon}} \right\},
   \label{eq:PSFest}
\end{equation}
where $O_{K^{w}_{p,q}}^{(k)} = \mathcal{F}\{o^{(k)} K^{w}_{p,q}\}$.

\subsection{PSF step in the spatial domain}
\label{:PSFSpatial}
The final step is the projection of the estimated local PSFs $\bar{h}_{p,q|s}^{(k+1)}$ on the feasible set of PSFs. All PSFs are constraint to be limited in size. Since morph occurs in the images, the local PSFs are unlikely to be centralized, therefore adaptive PSF support is used. Let $\mathbb{X}_{p,q|s}^{(k)}$ denote the adaptive PSF support for local PSF $h_{p,q|s}^{(k+1)}$. The adaptive PSF support is given by a circular area with radius, $r_{\mathbb{X}}$, and its centre is adaptive to the centre of mass of the local PSF. The set of all possible PSFs, $\mathcal{H} = \left\{h|h_{p,q|s}[m,n]=0 \text{ for } [m,n]\not\in\mathbb{X}_{p,q|s}^{(k)}\right\}$, is limited to the size of the support constraint. Furthermore, the PSF is an intensity distribution that contains only non-negative real values and is normalized:
\begin{equation}
    h_{p,q|s}^{(k+1)} = \frac{\max\left\{ \text{real}\left(\bar{h}_{p,q|s}^{(k+1)}\right),0 \right\}}{\left|\left|\max\left\{ \text{real}\left(\bar{h}_{p,q|s}^{(k+1)}\right) ,0\right\}\right|\right|_{1}}.
\end{equation}
As a result the PSFs can be written as the sum of two components. The tangential component of $h_{p,q|s}^{(k+1)}$ denoted as $\left(h_{p,q|s}^{(k+1)}\right)_{\parallel}$, is contained in $\mathcal{H}$. The normal component $\left(h_{p,q|s}^{(k+1)}\right)_{\perp}$ consists of the pixels of the PSF that are not contained in $\mathcal{H}$.
\begin{equation}
    h_{p,q|s}^{(k+1)} = \left(h_{p,q|s}^{(k+1)}\right)_{\parallel} + \left(h_{p,q|s}^{(k+1)}\right)_{\perp}, \quad \left(h_{p,q|s}^{(k+1)}\right)_{\parallel}\in\mathcal{H},\text{ }\left(h_{p,q|s}^{(k+1)}\right)_{\perp} = 0.
\end{equation}
Finally, the local PSF spectrum is given by the Discrete Fourier Transform of the estimated local PSF
\begin{equation}
    H_{p,q|s}^{(k+1)} = \mathcal{F} \left\{ h_{p,q|s}^{(k+1)} \right\}.
\end{equation}
In TIP4AW, these four steps are repeated consecutively. Algorithm \ref{alg1} shows an outline of TIP4AW.

\begin{algorithm}[tbh] 
	\caption{TIP4AW} 
	\label{alg1} 
	\begin{algorithmic} 
		\REQUIRE $\{i_1,...,i_S\}$, $k_{\text{max}}$, $r_{\mathbb{X}}$, $P$, $Q$, $w$, $\Delta w$, $\epsilon$, $p_s$
		\RETURN $\hat{o}$, $\{\hat{h}_{p,q|s}\}$
		\STATE initialize $h_{p,q|s}=\delta[\floor*{m/2},\floor*{n/2}]$
		\FOR{$k = 1,...,k_{\text{max}}$}
		\FOR{$p=1,...,P$\tikzmark{top1}}  
		\FOR{$q=1,...,Q$} 
		\STATE $\bar{o}_{p,q}^{(k)} = \mathcal{F}^{-1} \left\{ \frac{\sum_{s=1}^S a^{(k)}_{p,q|s} \left(H_{p,q|s}^{(k)}\right)^*I_{s}}{\left[\sum_{s=1}^S a^{(k)}_{p,q|s} |H_{p,q|s}^{(k)}|^2\right]_{>\epsilon\bar{a}_{p,q}}} \right\}$
		\ENDFOR
		\ENDFOR            \tikzmark{bottom1}
		\STATE $\bar{o}^{(k)} = \sum_{p=1}^P\sum_{q=1}^Q \bar{o}^{(k)}_{p,q}W_{p,q}$ \tikzmark{top2}
		\STATE $o^{(k)} = \frac{\max\left\{ \text{real}\left(\bar{o}^{(k)}\right),0 \right\}}{\left|\left|\max\left\{ \text{real}\left(\bar{o}^{(k)}\right) ,0\right\}\right|\right|_{1}}$    \tikzmark{bottom2}
		\STATE $O_{K^{w}_{p,q}}^{(k)} = \mathcal{F}\{o^{(k)} K^{w}_{p,q}\}$
		\STATE $O_{K^{w+\Delta w}_{p,q}}^{(k)} = \mathcal{F}\{o^{(k)} K^{w+\Delta w}_{p,q}\}$
		\FOR{$p=1,...,P$\tikzmark{top3}} 
		\FOR{$q=1,...,Q$}
		\FOR{$s=1,...,S$}
		\STATE $\bar{h}_{p,q|s}^{(k+1)} = \mathcal{F}^{-1} \left\{  \frac{I_s}{ \left[ O_{K^{w}_{p,q}}^{(k)} \right]_{>\epsilon}} \right\}$
		\STATE $\tilde{h}_{p,q|s}^{(k+1)} = \mathcal{F}^{-1} \left\{  \frac{I_s}{ \left[ O_{K^{w+\Delta w}_{p,q}}^{(k)} \right]_{>\epsilon}} \right\}$
		\ENDFOR
		\ENDFOR
		\ENDFOR \tikzmark{bottom3}
		\STATE $ h_{p,q|s}^{(k+1)} = \frac{\max\left\{ \text{real}\left(\bar{h}_{p,q|s}^{(k+1)}\right),0 \right\}}{\left|\left|\max\left\{ \text{real}\left(\bar{h}_{p,q|s}^{(k+1)}\right) ,0\right\}\right|\right|_{1}}$ \tikzmark{top4}
		\STATE $h_{p,q|s}^{(k+1)}\in\mathcal{H} = \left\{h|h_{p,q|s}[m,n]=0 \text{ for } [m,n]\not\in\mathbb{X}_{p,q|s}^{(k)}\right\}\qquad$  \tikzmark{right}\tikzmark{bottom4}
		\STATE $ a^{(k)}_{p,q|s} = ||h_{p,q|s}-\tilde{h}_{p,q|s}||_F^{-2p_s}$ 
		\STATE $H_{p,q|s}^{(k+1)} = \mathcal{F} \left\{ h_{p,q|s}^{(k+1)} \right\}$
		\ENDFOR
		\STATE $\hat{o} = o$ and $\hat{h}_{p,q|s} = h_{p,q|s}$
	\end{algorithmic}
	\AddNote{top1}{bottom1}{right}{Object step in Fourier domain (1)}
	\AddNote{top2}{bottom2}{right}{Object step in spatial domain (2)}
	\AddNote{top3}{bottom3}{right}{PSF step in Fourier domain (3)}
	\AddNote{top4}{bottom4}{right}{PSF step in spatial domain, for $h$ and $\tilde{h}$ (4)}
\end{algorithm}

\subsection{Weight determination}
\label{sec:weights}
They key novelty of the algorithm is weighted multi-frame deconvolution. For weighted multi-frame deconvolution the weights, $a_{p,q|s}$, represent the contribution of a frame, $i_s$, that is used in Eq.~\eqref{eq:WMFD} in order to estimate the object. If all weights are chosen $a_{p,q|s}=1$ and $\epsilon=0$ then Eq.~\eqref{eq:WMFD} is exactly the filter in the Fourier domain given in Eq.~\eqref{eq:Multi-Frame-Wiener}. Deconvolution in the Fourier domain requires the assumption that Eqs.~\eqref{eq:DiscIsoImageAndnoise} and \eqref{eq:OTF} are valid. This assumption only holds locally and it is possible that isoplanatic subsection do not coincide, especially when there is a lot of morph in the images. For optimal performance of the algorithms it is favorable that the image subsections are highly isoplanatic, that is with less variant blur. Weighted multi-frame deconvolution makes that highly isoplanatic image subsections contribute more to the deconvolution than image subsections that contain large anisoplanatic aberrations. 

When an image frame is locally highly isoplanatic the corresponding weight will be large and likewise a small weight will be assigned to a frame when it is locally anisoplanatic. The metric that is associated with the scale of the variation of the SV PSF is called \textit{isoplanatism}, or PSF variation. The process of finding optimal values for these weights has not been researched yet. 

In this paper a method is proposed to demonstrate the effect weighted multi-frame deconvolution has on the image quality of the estimated object. Isoplanatism is measured by comparing the similarly between the estimated local PSF, $h_{p,q|s}$, and complementary estimated local PSF, $\tilde{h}_{p,q|s}$. The local PSFs, as calculated in Eq.~\eqref{eq:PSFest}, depend on the width of the apodization kernel, $K^w$. For the calculation of the estimated local PSFs an apodization kernel, $K^w$, is used and for the complementary estimated local PSFs a wider apodization kernel, $K^{w+\Delta w}$, is used. Since the PSF is spatially variant, the complementary local PSF will be different than the local PSF. The difference is a measure for the local isoplanatism. The weights are given by the following formula:
\begin{equation}
    a_{p,q|s} = ||h_{p,q|s}-\tilde{h}_{p,q|s}||_F^{-2P_s},
    \label{eq:isoplanatism}
\end{equation}
where $||\cdot||_F$ denotes the Frobenius norm and $p_s$ is the isoplanatism sensitivity. This metric was first reported in the Master of Science thesis of one of the authors~\cite{Ketterij2019}.

\section{Results}
The performance of TIP4AW is compared with the space-variant Online Blind Deconvolution with Efficient Filter Flow (OBD/EFF) algorithm in~\cite{Hirsch2010}. This algorithm was implemented in the best of the author's ability. Thereafter the robustness to noise of both algorithms is compared. A MATLAB implementation of TIP4AW and a set of test images can be found here \cite{TIP4AW}.

\subsection{Algorithm comparison}
Before both algorithms can be compared, it is necessary to define an adequate quality metric. As for simulated data the estimated object can be compared to the real object or ground truth. Due to large morph in the images the authors do not recommended to use an error based metric such as the PSNR or mean squared error.

Instead a metric called Fourier Ring Correlation (FRC) which measures the normalized cross-correlation between the real object and the estimated object on rings, $r_n$, in the Fourier domain was used. FRC is originally a method to determine the resolution of reconstructed images in electron microscopy~\cite{frc} and later used in~\cite{banterle} as a resolution criterion in super-resolution microscopy. In this thesis FRC will be used to measure the similarity between the real object and the estimated object,
\begin{equation}
    FRC[r] = \frac{\sum_{r_n\in r}O[r_n]\cdot \hat{O}^*[r_n] }{\sqrt{\sum_{r_n\in r}\left|O[r_n]\right|^2\cdot\sum_{r_n\in r}\left|\hat{O}[r_n]\right|^2}}.
\end{equation}
An FRC curve close to one corresponds to a close match of the real object and estimated object. Since the algorithms are independent of the imaging system, the resolution of the imaging sensor is unknown. The quality of the estimated object can be expressed by a single value that is the largest ring, $r_{n,\text{max}}$, where the FRC was larger than a reference line, the ${2\sigma}$ curve, used in both~\cite{frc,banterle}.

\begin{figure}[tbh]
    \centering
    \includegraphics[width=\textwidth]{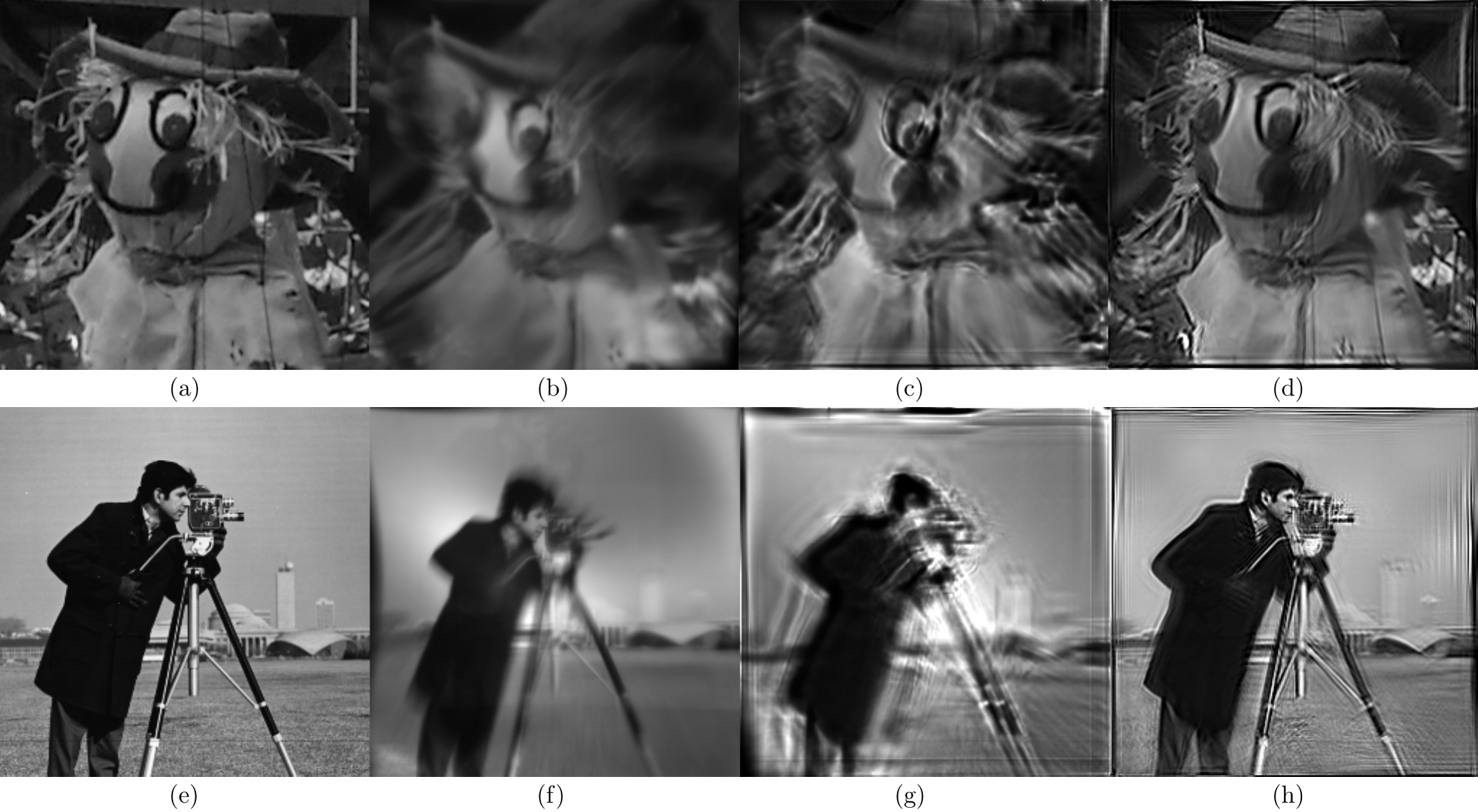}
    \caption{(a)~The ground truth \textit{Puppet}, (b)~a typical image frame of Puppet, (c)~estimated object (Puppet) by OBD/EFF and (d)~estimated object (Puppet) by TIP4AW. (e)~The ground truth \textit{Cameraman}, (f)~a typical image frame of Cameraman, (g)~estimated object (Cameraman) by OBD/EFF and (h)~estimated object (Cameraman) by TIP4AW. OBD/EFF parameters are: $P=Q=7$, PSF size: 13 pixels, SRF: 2. TIP4AW parameters are: $P=Q=7$, $r_{\mathbb{X}}=6$ (13 pixels), $w=35$, $\Delta w=14$, $\epsilon=10^{-4.4}$, $p_s=1.5$. Quantitative results are shown in Fig.~\ref{fig:FRCCompareHirsch}.}
    \label{fig:CompareHirsch}
\end{figure}

Two numerical experiments were conducted wherein the object was estimated by TIP4AW and OBD/EFF. In each experiment 30 images of the same object were provided to the algorithms. For the experiments the size of the PSF was estimated at 13 pixels and the images are divided into 49 overlapping subsection. For TIP4AW the apodization kernels, $K^{35}$ and $K^{49}$, noise parameter, $\epsilon = 10^{-4.4}$, and isoplanatism sensitivity, $P_s=1.5$, were used in order to maximize the performance. For OBD/EFF a super resolution factor~(SRF) of 2 was used. OBD/EFF uses one image in every iteration, therefore 30 iterations are chosen for TIP4AW. Figure~\ref{fig:CompareHirsch}(a)-(h) shows the ground truth, a typical image frame, and the estimated object for both OBD/EFF and TIP4AW.

\begin{figure}[!ht]%
\centering
    \subfloat[]{{\includegraphics[width=\textwidth,valign=c]{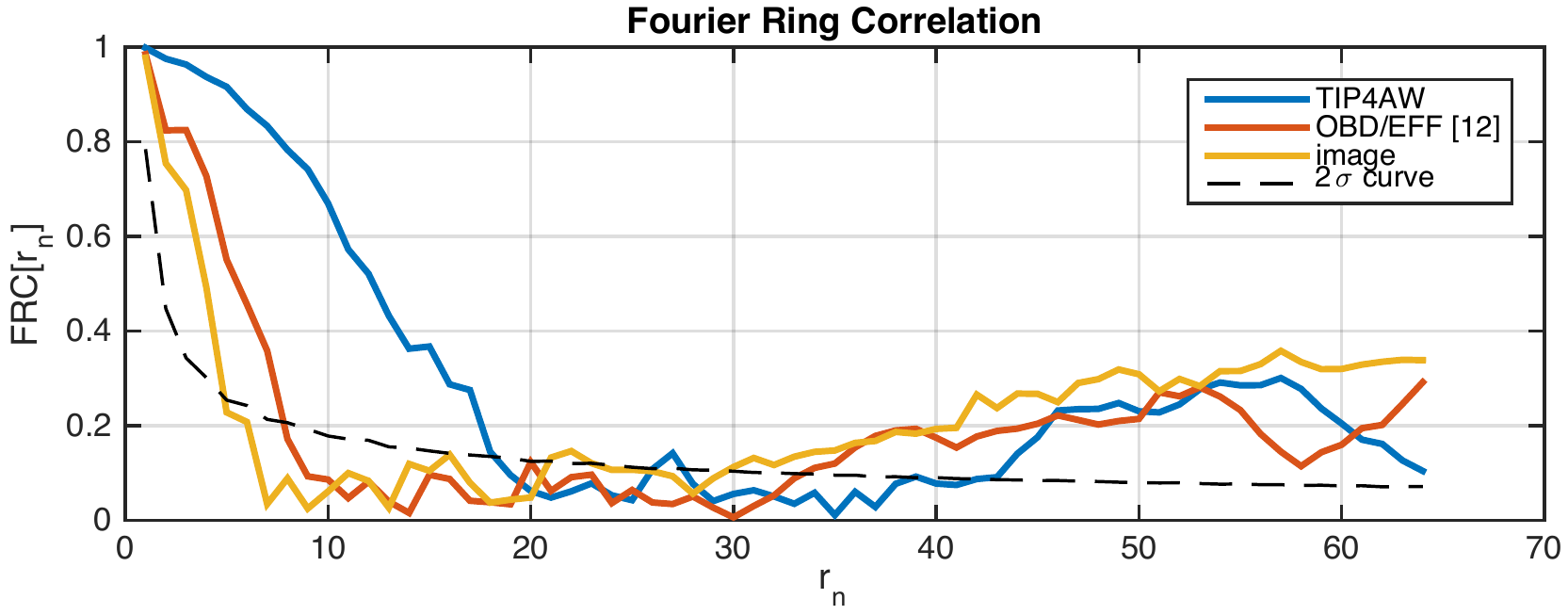} }}%
    \vspace{0.1cm}
    \subfloat[]{{\includegraphics[width=\textwidth,valign=c]{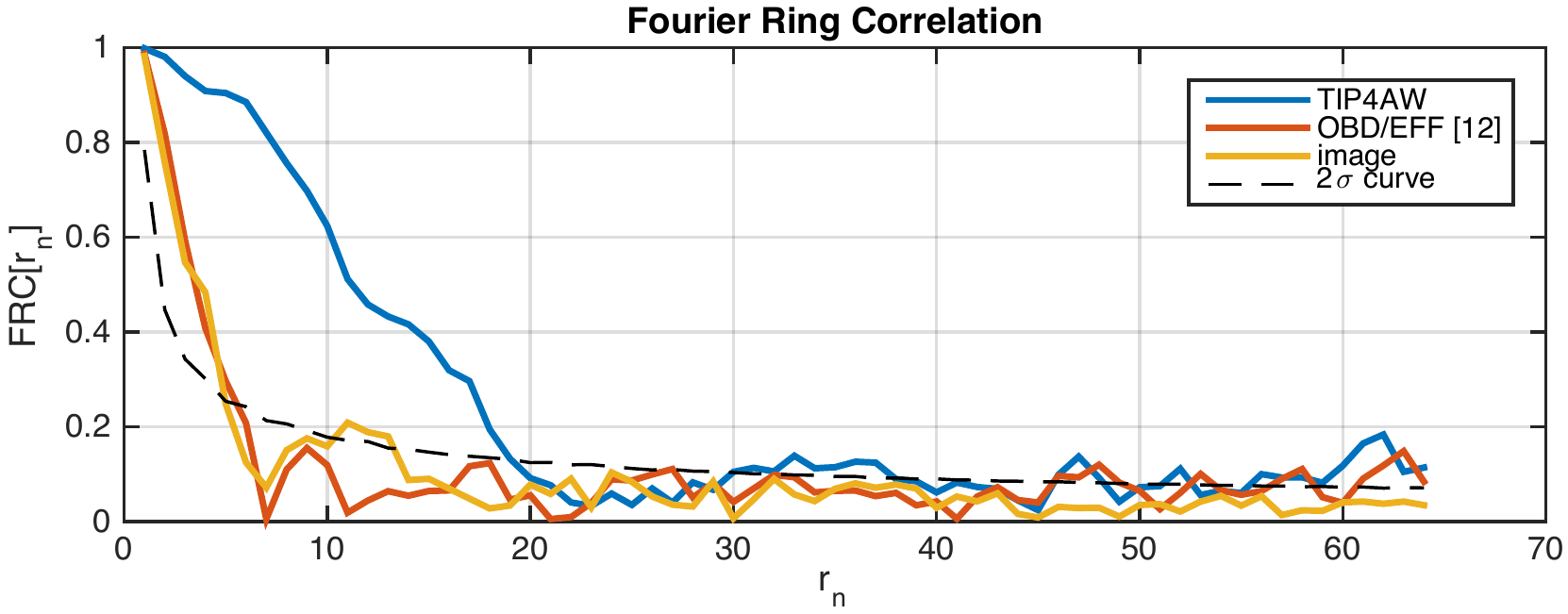} }}%
    \caption{The Fourier Ring Correlation provides a quantitative comparison of the results depicted in Fig.~\ref{fig:CompareHirsch}. The quality of the estimated object is compared to the ground truth and can be expressed as the crossing of the FRC and the $2\sigma$ curve. This crossing is denoted by the largest Fourier ring, $r_{n,\text{max}}$. (a) The quality of the image frame, estimated object with OBD/EFF and TIP4AW in Fig.~\ref{fig:CompareHirsch}(b)-(d) are $r_{n,\text{max}}=4$, $7$ and $18$ respectively.
    (b) The quality of the image frame, estimated object with OBD/EFF and TIP4AW in Fig.~\ref{fig:CompareHirsch}(f)-(h) are $r_{n,\text{max}}=4$, $5$ and $18$ respectively.
    }%
    \label{fig:FRCCompareHirsch}%
\end{figure}

Figure~\ref{fig:FRCCompareHirsch} shows a quantitative comparison by the FRC corresponding to the estimated object for both OBD/EFF and TIP4AW. For reference the FRC of the image frames in Fig.~\ref{fig:CompareHirsch}, and the $2\sigma$ curve are shown.

\subsection{Effect of additive noise}
In the experiments with TIP4AW and OBD/EFF in this section, Gaussian white noise was added to the images. The pixels in the images have a range from $0$ to $1$ and the Gaussian noise has varying standard deviation. For every noise level the experiments were repeated 50 times. The quality of the estimated object is measured with the Structural Similarity Index (SSIM). SSIM is a metric inspired be the human visual system~\cite{SSIM}. SSIM is a method designed to measure the visual similarity of two images. Here SSIM is used to measure the influence of additive noise. SSIM is a measurement between $0$ and $1$, where values close to $1$ stand for high similarity.
\begin{equation}
SSIM = \frac{ (2\mu_o\mu_{\hat{o}}+c_1) (2\sigma_{o\hat{o}}+c_2) }
{ (\mu_o^2+\mu_{\hat{o}}^2+c_1) (\sigma_{o}^2+\sigma_{\hat{o}}^2+c_2) },
\end{equation}
where $\mu_o$, $\mu_{\hat{o}}$, $\sigma_o$, $\sigma_{\hat{o}}$ and $\sigma_{o\hat{o}}$ are the local means, standard deviations and cross-covariance for the object and estimated object respectively. $c_1$ and $c_2$ are constants. 

\begin{figure}[tbh]
    \centering
    \includegraphics[width=\textwidth]{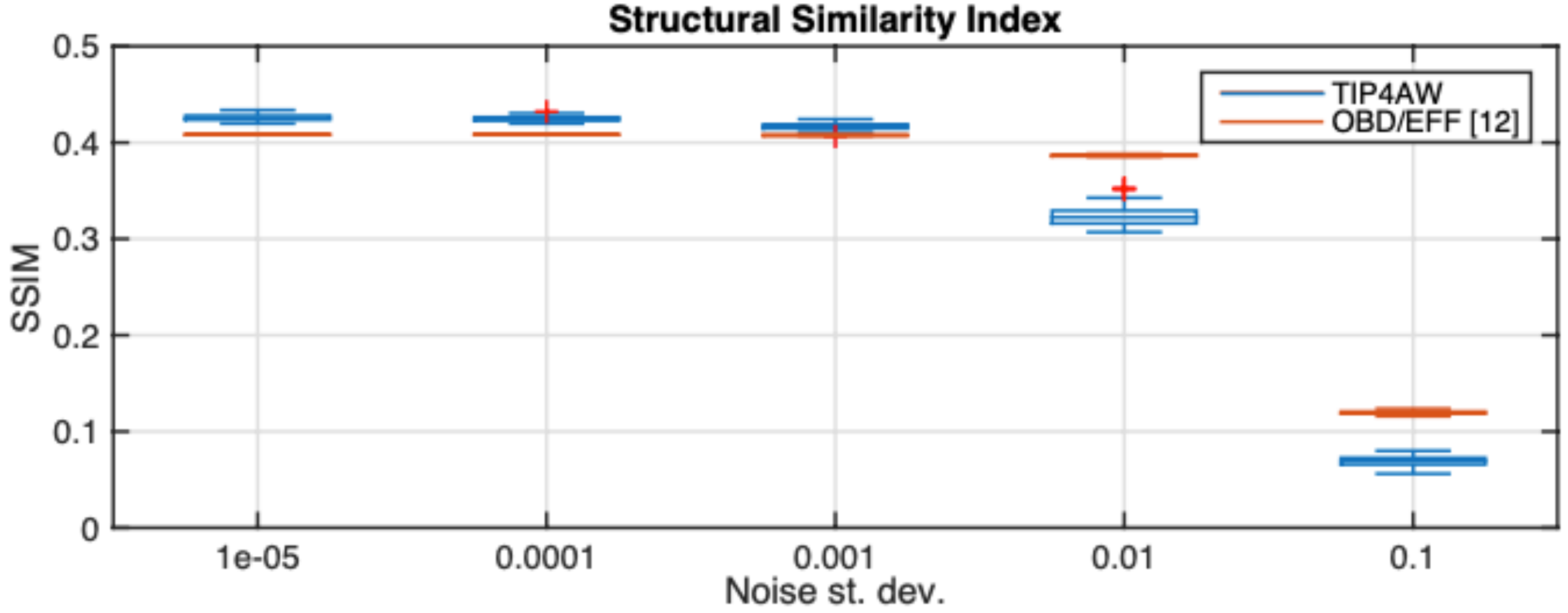}
    \caption{Image quality of the estimated object with TIP4AW and OBD/EFF. White Gaussian noise with different standard deviations (st. dev.) was added to the set of images. The image quality is measured with SSIM. Qualitative results are shown in Fig.~\ref{fig:Noise}.}
    \label{fig:NoiseBox}
\end{figure}

Figure~\ref{fig:NoiseBox} shows the image quality of the estimated object for different noise levels. The standard deviations of the noise was varying form $\sigma=10^{-5}$ to $\sigma=10^{-1}$.

\begin{figure}[tbh]
    \centering
    \includegraphics[width=\textwidth]{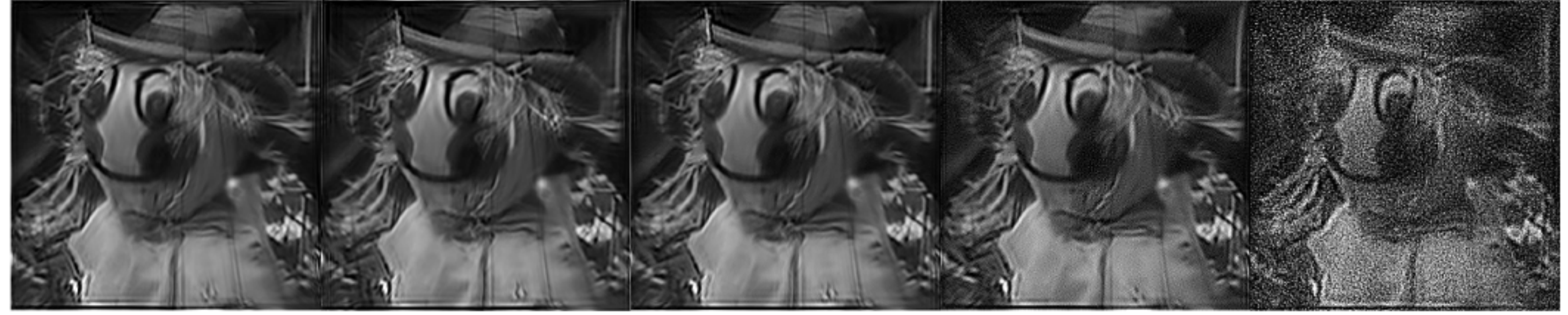}
    \caption{In an experiment TIP4AW is used to estimate the original object from sets of input images with different noise levels. White Gaussian noise with varying standard deviations, $\sigma=10^{-5}, 10^{-4}, 10^{-3}, 10^{-2}, 10^{-1}$, is added to the images. The estimated object corresponding to the sets of input images with different noise levels are depicted in this figure respectively. The quality in SSIM of the estimated objects can be found in Fig.~\ref{fig:NoiseBox}.}
    \label{fig:Noise}
\end{figure}

The estimated objects with TIP4AW for a set of input images with different standard deviation are shown in Fig.~\ref{fig:Noise}. The images from left to right are the result of TIP4AW with input images with addition white Gaussian noise with increasing standard deviation. 

\section{Discussion and conclusion}
\label{::Discussion}
The term deconvolution refers the inverse process of convolution. Since convolution for image formulation can only be assumed for isoplanatic aberrations, the term anisoplanatic deconvolution is abused in literature when anisoplanatic aberrations are considered. Yet, this paper refers to deconvolution. In this paper weighted multi-frame deconvolution is used to deconvolve multiple frames in order to estimate the object from blurred images. If local isoplanatism in images with anisoplanatic aberrations is sufficient, the performance of TIP4AW can reach what was possible with a batch of images with only isoplanatic aberrations.

TIP4AW was compared to OBD/EFF~\cite{Hirsch2010}. Two data sets, \textit{Puppet} and \textit{Cameraman}, were used and both algorithms were used to calculate the underlying object. Figure.~\ref{fig:CompareHirsch} shows the qualitative results of the experiments and Fig.~\ref{fig:FRCCompareHirsch} show that the estimated object with TIP4AW is of higher quality than OBD/EFF for these two data sets and in terms of FRC. Especially in the Fourier rings corresponding to low spatial frequencies TIP4AW is superior. It is hypothesised that online algorithms that process one image per iteration are not capable of handling large morph in images. TIP4AW is specifically designed for situations where large morph in the images occurs. For this the estimated object with TIP4AW will show a good performance for low spatial frequencies. For the best performance of TIP4AW it is recommended to obtain as much images of the same object as possible. It is not required to align the images, as the algorithm can correct for translations in the object by itself. Obtaining more images will result in a better image quality because it is more likely for isoplanatic subsections to occur and to measure more frequencies within the bandlimit of the system. As demonstrated in~\cite{Wilding2017} and~\cite{Ketterij2019}, TIP-based algorithm have limited performance when the object is sparse, for example in stellar imaging.

In sparse parts of the object the frequency content is not sufficient for accurate estimation of the local PSF. Consequently ringing occurs at the edge of a restored image. In Fig.~2(h) the effect of ringing is noticeable along and normal to the edges of the image. Since the local PSF depends on the image and the object as in Eq.~\eqref{eq:PSFest}, and due to the anisoplanatic nature of the aberration, it is that the effect of ringing occurs non-uniformly. A possible solution for this problem is the calculation of local PSFs at the edges of the images or the use of edge enhancement software.

Besides ringing, shadowing occurs in the restored images. Comparing Fig.~2(d) with ground truth, Fig.~2(a), it can be seen that a shadow has appeared left of the puppet's eye. Also, a shadow occurs at the left of the cameraman's coat, Fig.~2(h), that was not visible in the ground truth, Fig.~2(e). The shadow effect occurs when the estimated local PSF is larger than the adaptive support area. The size of the adaptive support area is a trade-off between accuracy of the estimated local PSF and the convergence of the algorithm.

TIP is known for its robustness to noise~\cite{Wilding2017} and the same holds for TIP4AW. In Figs.~\ref{fig:NoiseBox} and~\ref{fig:Noise} is demonstrated that both TIP4AW and OBD/EFF are robust to noise. The quality of the estimated object after 50 repetitions of both algorithms for different noise levels was measured by the SSIM. As can be seen in Fig.~\ref{fig:NoiseBox}, for low noise levels ($10^{-5}$) the quality of the estimated object with TIP4AW is higher than OBD/EFF in terms of SSIM. As the noise standard deviation increases the quality of the estimated object decreases. Figure~\ref{fig:NoiseBox} shows that the quality with TIP4AW decreases more with increasing noise than OBD/EFF. Therefore, it can be concluded that OBD/EFF is more robust to noise than TIP4AW. Nevertheless, even in noisy situations the quality of the estimated object with TIP4AW will be higher in terms of FRC.

\begin{figure}[tb]
    \centering
    \includegraphics[width=\textwidth]{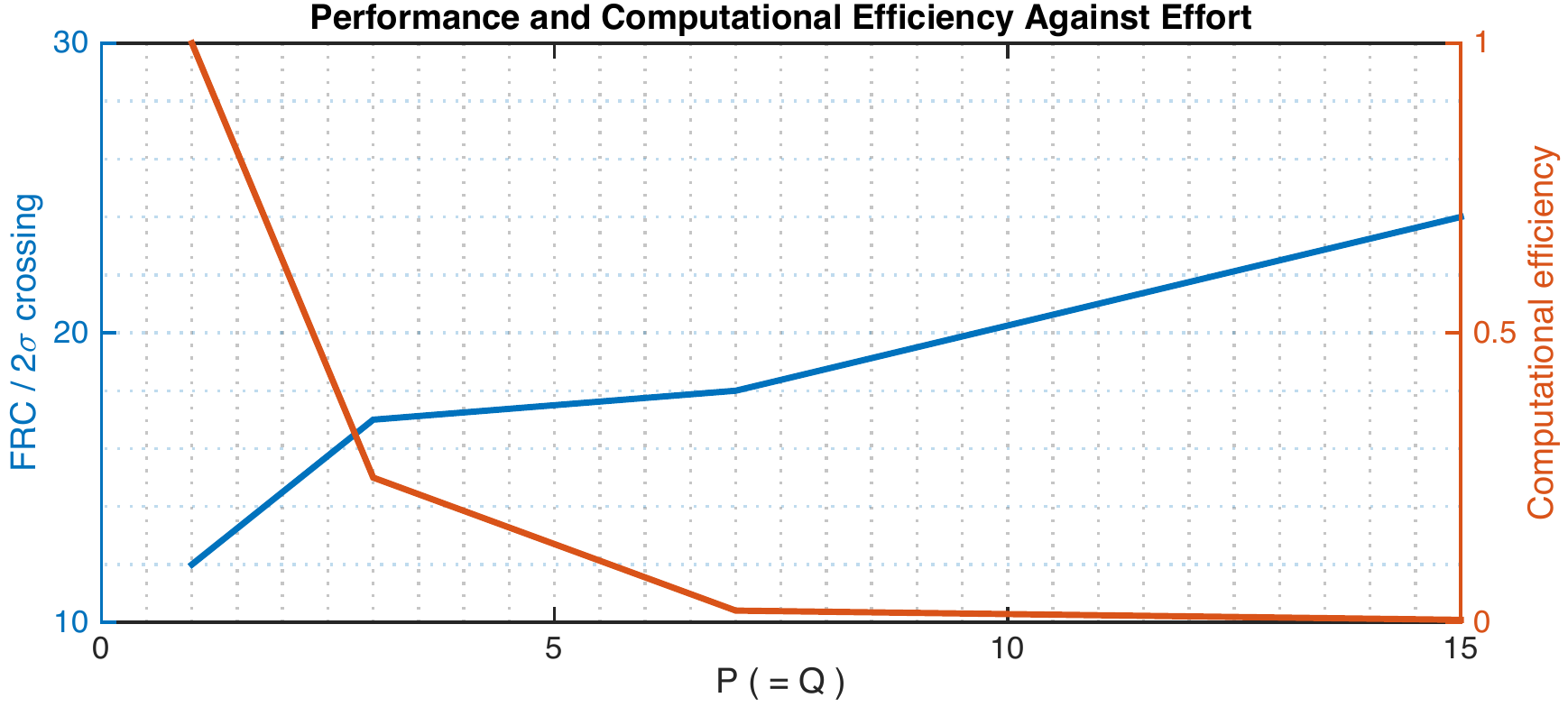}
    \caption{The performance of TIP4AW given by the crossing of the FRC with the $2\sigma$ curve increases with the the number of subsections, $PQ$. In this example the horizontal and vertical number of subsections is equal, $P=Q$. The computational efficiency is given by $1/(PQ)$ and decreases with an increasing number of subsections. As expected the performance of TIP4AW increases as the number of subsections increases.}
    \label{fig:PerformanceVsCost}
\end{figure}

The performance of TIP4AW does not only depend on the number of images and the level of noise. Also the parameters of the algorithm have effect. In general the algorithm requires some iterations to converge to a solution, typically 20 to 30 iterations is sufficient. Further, the user should have a guess about the size of the PSF in order to set the size of the PSF support. If the user has no information or insight in the size of the PSF the algorithm can be run multiple times for different support constraints. The size of the Gaussian apodization kernel is such that its full width at half maximum is about 2 to 4 times larger than the PSF support. The apodization size should be as small as possible while there is still sufficient information for the local PSF to be estimated, therefore the width of the Gaussian apodization kernel depends on the object and aberration. The parameter $\epsilon$ is related to the SNR and is usually small. Moreover, the quality of the estimated object will increase if more subsections are used. The computational effort of TIP4AW is mainly determined by the number of subsections, $PQ$. Figure~\ref{fig:PerformanceVsCost} shows the relation between the performance of TIP4AW measured by the crossing of the FRC and the $2\sigma$ curve and the computational efficiency, $1/(PQ)$. 

Furthermore, an influence on the quality of the estimated object is the initial aberration in the recorded images. The anisoplanatic effects will always remain in the corrected image. This also holds for other image processing algorithms, unless a correction is made for every pixel independently. It is the strength of TIP4AW, that when some (nearly) isoplanatic regions occurs, this desired region will have a larger contribution to the final result. Therefore, TIP4AW actively suppresses the anisoplanatic effects.

In this paper was chosen for an approach where the least amount of prior knowledge of image formation was assumed. Note that TIP4AW works exclusively with local PSFs and not with information about the unknown phase aberration itself. Only two \textit{a priori} are assumed for the local PSFs being non-negative and real-valued, and limited in size. Both assumptions are applied in the form of a spatial domain constraint to the PSF. The advantage of this approach is that the shape of the PSF is not limited to a parameterized model. The object has a non-negative and real-valued constraint.

This paper demonstrates the potential of weighted multi-frame deconvolution. This novel approach to deconvolution was performed in conjunction with PSF variation or isoplanatism. The calculation of the weights $a_{p,q|s}$ as conducted in this paper is not optimal. Further research should be done about optimal use of weighted multi-frame deconvolution. A possible use for weighted multi-frame deconvolution is known as lucky imaging~\cite{Fried78}. When not the isoplanatism but the size of the PSF is measured, for example by the second moment of the estimated PSF, it is possible to use TIP4AW as a lucky imaging algorithm. In lucky imaging non degraded regions in the image are used for reconstruction of the object.

Another approach in this research could make explicit use of the phase aberrations. Assuming a single conjugation plane outside the pupil plane leads to a phase retrieval problem. Explicit use of the anisoplanatic phase aberration might lead to more coherence between the local PSFs.

To conclude, TIP4AW was presented as an algorithm that can handle with anisoplanatic aberrations and is robust to noise and morph. This algorithm requires multiple images of the same object and only a guess about the size of the PSF. TIP4AW is a practical algorithm for situations where space-temporal variations of the aberration is present. In general TIP4AW is an algorithm that works in batch mode, where all images are processed at the same time. Most of the computational time is spend during the PSF step in the Fourier Domain that is step (3) in algorithm~\ref{alg1}. Most of this computational time can be reduced by calculating all of the local PSFs in parallel. Besides, TIP4AW can also be implemented as an online algorithm. For online computation of TIP4AW, $S'$ images should be processed at the same time, where $S'<S$. In the first iteration images $\{i_1,...,i_{s'}\}$ are processed, in the second iteration the images $\{i_2,...,i_{s'+1}\}$, and in iteration $k$ the images $\{i_k,...,i_{s'+k-1}\}$ are processed. The use of a moving window of $S'$ images allows online implementation of TIP4AW.

\section*{Funding}
This project has received funding from the ECSEL Joint Undertaking (JU) under grant agreement No. 826589. 
The JU receives support from the European Union's Horizon 2020 research and innovation programme and Netherlands, Belgium, Germany, France, Italy, Austria, Hungary, Romania, Sweden and Israel.


\section*{Disclosures}
%
%
%
%
%

The authors declare no conflicts of interest.

\bibliography{sample}

\end{document}